\documentclass{knawproc}

\def\deg{\ifmmode ^\circ                
         \else $^\circ$
         \fi
         \hskip -0.1truecm}
\def\degd#1.#2{                         
               \ifmmode {#1^{\hskip 0.05em\circ}\hskip-0.42em.\hskip0.08em#2}
               \else {#1$^{\hskip 0.05em\circ}\hskip-0.42em.\hskip0.08em$#2}
               \fi
              }
\def\mind#1.#2{                         
               \ifmmode {#1^{\hskip 0.05em\prime}\hskip-0.35em.\hskip0.05em#2}
               \else {#1$^{\hskip 0.05em\prime}\hskip-0.35em.\hskip0.05em$#2}
               \fi
              }
\def\secd#1.#2{                         
               \ifmmode {#1^{\prime\prime}\hskip-0.46em.\hskip0.12em#2}
               \else {#1$^{\prime\prime}\hskip-0.46em.\hskip0.12em$#2}
               \fi
              }
\def\timsecd#1.#2{                      
                  \ifmmode {#1^{\rm s}\hskip-0.39em.\hskip0.08em#2}
                  \else {$#1^{\rm s}\hskip-0.39em.\hskip0.08em#2$}
                  \fi
                 }
\def\hms#1h#2m#3s{                      
                  \relax
                  \ifmmode #1^{\rm h}\,#2^{\rm m}\,#3^{\rm s}
                  \else \hbox{$#1^{\rm h}\,#2^{\rm m}\,#3^{\rm s}$}
                  \fi
                 }
\def\dms#1d#2m#3s{                      
                  \relax
                  \ifmmode #1^\circ\,#2^{\prime}\,#3^{\prime\prime}
                  \else \hbox{$#1^\circ\,#2^{\prime}\,#3^{\prime\prime}$}
                  \fi
                 }
\def\dmsd#1d#2m#3.#4s{                  
                      \relax
                      \ifmmode #1^\circ\,#2^{\prime}\,#3^{\prime\prime}
                               \hskip-0.46em.\hskip0.12em#4
                      \else \hbox{$#1^\circ\,#2^{\prime}\,#3^{\prime\prime}
                            \hskip-0.46em.\hskip0.12em#4$}
                      \fi
                     }
\def\hm#1h#2m{                          
              \relax
              \ifmmode #1^{rm h}\,#2^{\rm m}
              \else \hbox{$#1^{\rm h}\,#2^{\rm m}$}
              \fi
             }
\def\dm#1d#2m{                          
              \relax
              \ifmmode #1^\circ\,#2^{\prime}
              \else \hbox{$#1^\circ\,#2^{\prime}$}
              \fi
             }
\def\hmsd#1h#2m#3.#4s{                  
                      \relax
                      \ifmmode #1^{\rm h}\,#2^{\rm m}\,#3^{\rm s}
                               \hskip-0.39em.\hskip0.08em#4
                      \else \hbox{$#1^{\rm h}\,#2^{\rm m}\,#3^{\rm s}
                            \hskip-0.39em.\hskip0.08em#4$}
                      \fi
                     }
\def\hmd#1h#2.#3m{                  
                  \relax
                  \ifmmode #1^{\rm h}\,#2^{\rm m}
                           \hskip-0.55em.\hskip0.22em#3
                  \else \hbox{$#1^{\rm h}\,#2^{\rm m}
                        \hskip-0.55em.\hskip0.22em#3$}
                  \fi
                 }
\def\mg{\relax                          
        \ifmmode ^{\rm m}
        \else $^{\rm m}$
        \fi
       }
\def\mgd#1.#2{                          
              \relax
              \ifmmode #1^{\rm m}
                       \hskip-0.55em.\hskip0.22em#2
              \else \hbox{#1$^{\rm m}
                    \hskip-0.55em.\hskip0.22em$#2}
              \fi
             }

\def\la{\mathrel{\hbox{\rlap{\hbox{\lower4pt\hbox{$\sim$}}}\hbox{$<$}}}}
\def\ga{\mathrel{\hbox{\rlap{\hbox{\lower4pt\hbox{$\sim$}}}\hbox{$>$}}}}

\def\unitspace{\,}                      

\def\un#1{\ifmmode \unitspace{\rm #1} 
          \else $\unitspace$#1
          \fi}
\def\pun#1#2{\ifmmode \unitspace\mbox{\rm #1}^{#2} 
             \else $\unitspace$#1$^{#2}$
             \fi}

\def\kms{\un{km}\pun{s}{-1}}          
\def\Lsun{\ifmmode \un{L}_{\odot}     
          \else $\un{L}_{\odot}$
          \fi}
\def\Msun{\ifmmode \un{M}_{\odot}     
          \else $\un{M}_{\odot}$
          \fi}
\def\mum{\ifmmode \unitspace\mu\mbox{\rm m} 
         \else $\unitspace\mu$m
         \fi}

\def\Bp{\relax                            
        \ifmmode B_{||}                   
        \else $B_{||}$
        \fi}
\def\Bt{\relax                            
        \ifmmode B\!_{\perp}              
        \else $B\!_{\perp}$               
        \fi}
\def\Gcr{\relax                           
         \ifmmode \Gamma\!_{\rm cr}       
         \else $\Gamma\!_{\rm cr}$
         \fi}
\def\ICII{\relax                          
          \ifmmode I_{[\CII]}             
          \else $I_{[\CII]}$
          \fi}
\def\LHtwo{\relax                                 
           \ifmmode L_{\mbox{\rm\scriptsize H}_2} 
           \else $L_{\mbox{\rm\scriptsize H}_2}$  
           \fi}
\def\LLya{\relax                          
          \ifmmode L_{{\rm Ly}\,\alpha}   
          \else $L_{{\rm Ly}\,\alpha}$
          \fi}
\def\MHtwo{\relax                                 
           \ifmmode M_{\mbox{\rm\scriptsize H}_2} 
           \else $M_{\mbox{\rm\scriptsize H}_2}$  
           \fi}
\def\MHtwodot{\relax                                       
              \ifmmode \dot{M}_{\mbox{\rm\scriptsize H}_2} 
              \else $\dot{M}_{\mbox{\rm\scriptsize H}_2}$  
              \fi}                                         
\def\Mstardot{\relax                      
              \ifmmode \dot{M}_{\ast}     
              \else $\dot{M}_{\ast}$      
              \fi}
\def\nHI{\relax                                      
         \ifmmode n_{\mbox{\scriptsize\rm H\,\sc I}} 
         \else $n_{\mbox{\scriptsize\rm H\,\sc I}}$
         \fi}
\def\nHtwo{\relax                                
           \ifmmode n_{{\mbox{\scriptsize H}}_2} 
           \else $n_{{\mbox{\scriptsize H}}_2}$  
           \fi}

\def\sou#1#2{\relax                       
             \ifmmode {\rm #1}\,{\rm #2}  
             \else #1$\,$#2
             \fi}

\def\qu#1#2{\relax                          
            \ifmmode #1_{\rm #2}            
            \else $#1_{\rm #2}$
            \fi}

\def\mbox{\hbox}           

\def\CO#1{\ifnum#1=0                    
           \ifmmode \mbox{\rm CO}
           \else {\rm CO}
           \fi
          \else
           \ifnum#1<15
            \ifmmode ^{#1}\mbox{\rm CO}
            \else $^{#1}${\rm CO}
            \fi
           \else
            \ifmmode \mbox{\rm C}^{#1}\mbox{\rm O}
            \else {\rm C}$^{#1}${\rm O}
            \fi
           \fi
          \fi}

\def\COp{\ifmmode \mbox{\rm CO}^+           
         \else {\rm CO}$^+$                 
         \fi}

\def\CS#1{\ifnum#1=0                    
           \ifmmode \mbox{\rm CS}
           \else {\rm CS}
           \fi
          \else
           \ifnum#1<15
            \ifmmode ^{#1}\mbox{\rm CS}
            \else $^{#1}${\rm CS}
            \fi
           \else
            \ifmmode \mbox{\rm C}^{#1}\mbox{\rm S}
            \else {\rm C}$^{#1}${\rm S}
            \fi
           \fi
          \fi}

\def\HCOp{\ifmmode \mbox{\rm HCO}^+          
          \else {\rm HCO}$^+$                
          \fi}
\def\Hthreep{\ifmmode \mbox{\rm H}_3^+         
             \else {\rm H}$_3^+$               
             \fi}

\def\Htwo{\ifmmode \mbox{\rm H}_2              
          \else {\rm H}$_2$                    
          \fi}

\def\HtwoO{\ifmmode \mbox{\rm H}_2\mbox{\rm O} 
           \else {\rm H}$_2${\rm O}            
           \fi}

\def\ion#1#2{\ifmmode \mbox{{\rm #1}}\,\mbox{{\sc #2}} 
        \else {\rm #1}$\,${\sc #2}
        \fi}

\def\rec#1#2{\if#2a                            
              \ifmmode \mbox{{\rm #1}}\alpha   
              \else {\rm #1}$\alpha$
              \fi
             \fi
             \if#2b
              \ifmmode \mbox{{\rm #1}}\beta
              \else {\rm #1}$\beta$
              \fi
             \fi
             \if#2g
              \ifmmode \mbox{{\rm #1}}\gamma
              \else {\rm #1}$\gamma$
              \fi
             \fi}

\def\Lya{\rec{Ly}{a}}                          

\newcommand{\tabref}[1]{Table~\protect\ref{#1}}
\newcommand{\figref}[1]{Fig.~\protect\ref{#1}}

\newcommand{\eqref}[1]{Eq.~$\left(\protect\ref{#1}\right)$}
\newcommand{\eqsref}[2]{Eqs.~$\left(\protect\ref{#1}\right)$--$\left(\protect\ref{#2}\right)$}
\newcommand{\secref}[1]{Sect.~\protect\ref{#1}}

\newcommand{\Tabref}[1]{\tabref{#1}}
\newcommand{\Figref}[1]{Figure~\protect\ref{#1}}

\usepackage{pvdwbib}
\usepackage{psfig}

\begin{document}

\begin{opening}

\title{\vspace*{-1cm}Dust and molecular gas in high redshift radio galaxies}

\author{Paul P.\ van der Werf}
\addresses{Leiden Observatory,
           P.O.~Box 9513,
           NL--2300 RA Leiden,
           The Netherlands
           (pvdwerf@strw.leidenuniv.nl)\\
}

\runningtitle{Dust and molecular gas in high-$z$ radio galaxies}
\runningauthor{Paul P.\ van der Werf}

\end{opening}

\begin{abstract}
This review discusses the current status of our
knowledge of emission by dust and molecular gas in high redshift radio
galaxies, and the uncertainties in the derivation of physical parameters
from these data. The similarity of far-infrared luminous,
gas-rich high redshift radio galaxies and local
ultraluminous infrared galaxies (ULIGs) is discussed. Given that local
ULIGs rapidly convert most of their gas reservoir into stars,
far-infrared luminous high-$z$ radio galaxies are likely undergoing
immense bursts of star formation, possibly accounting for a large
fraction of the final stellar populations in these systems. 
These results are discussed in the context 
of formation scenarios of massive galaxies.
\end{abstract}

\renewcommand{\thefootnote}{}

\section{Introduction}

In recent years, evidence has been 
accumulating that radio galaxies at 
various\footnote{\hspace{-1.8em}{\sl to appear 
in {\it The most distant radio galaxies},
    eds. H.J.A.\ R\"ottgering, P.N.\ Best, \& M.D.\ Lehnert, 
    Kluwer Academic Publishers (Dordrecht)}}
redshifts can be used to trace the formation and evolution of
massive elliptical galaxies. At low redshift, it has long been known that
that
radio-loud active galactic nuclei (AGNs) are found predominantly in luminous
($L\sim2-5\,L_*$) elliptical galaxies, which are often the dominant cD
galaxies of rich clusters (e.g., \citebare{Matthewsetal64}). 
At $z\sim1$, the host galaxies of well-studied
3CR sources are fully formed massive ellipticals
\cite{Bestetal98a}. High redshift ($z>2$) radio galaxies on the other
hand, show a variety of morphologies, which is often characterized by
the presence of a number of clumps (e.g., \citebare{Mileyetal92};
\citebare{Pentericcietal97}), totally unlike the more relaxed appearance
of their lower $z$ counterparts. The properties of these high-$z$
radio galaxies likely reflect processes related directly to the
buildup of the bulk of the stellar population. Hence the formation of
present-day massive ellipticals may be observed directly in high-$z$
radio galaxies.

The purpose of this review is to address the implications of the
recent detections of dust emission in a number of
high-$z$ radio galaxies for our understanding of the role of these
objects in the formation and early evolution of massive
galaxies. After a description of the general observational and
theoretical context (\secref{sec.context}), the present observational
situation in this rapidly developing field is reviewed
(\secref{sec.obs}). These results are then discussed in the
context of what we know about far-IR luminous galaxies in the local universe
(\secref{sec.local}).
Throughout this paper we adopt a value $H_0=75\,\kms\,\pun{Mpc}{-1}$
for the Hubble constant, and a deceleration parameter $q_0=0.1$,
unless otherwise indicated. Where necessary, results from other
publications are tacitly converted to the cosmology adopted here.

\section{Formation scenarios of massive galaxies}\label{sec.context}

It is convenient for the purpose of this review to divide galaxy
formation scenarios into two types.
\begin{enumerate}
\item {\it Hierarchical clustering\/} scenarios emphasize the gradual
buildup of massive objects from initially smaller fragments as the
main driver of the evolution of the galaxy population (e.g.,
\citebare{WhiteFrenk91}; \citebare{Kauffmannetal93}; \citebare{Baughetal98}).
\item The {\it monolithic collapse\/} scenario is based on the
idea that an {\it initial starburst\/} occurs, in which a galaxy forms
the bulk of its stellar population in a short amount of time; during
this starburst, most of the baryonic mass of the galaxy is turned from
gas into stars.
\end{enumerate}

It should be emphasized from the outset, that these two scenarios are
not as diametrically opposed as they seem. For instance, in the
hierarchical
clustering scenario, a major merger is required to form an elliptical
galaxy. In the local universe, such major mergers are known to be
accompanied by intense starbursts, as observed in the ultraluminous
infrared galaxies (ULIGs), in which a large fraction of the available
gas reservoir is rapidly converted into stars.
Therefore, even though 
they are not currently ``forming'', the local ULIGs possess 
all the properties of a massive galaxy building up its stellar population
by monolithic collapse. Thus even within the framework of 
hierarchical merging, galaxies that behave {\it as if\/} 
they are undergoing monolithic 
collapse exist.

Since the monolithic collapse scenario would result in a galaxy,
the stellar component of which is dominated by an evolved population, 
this scenario has been applied mostly to the formation of
spheroids.
The recent discoveries of luminous (up to a few times $10^{14}\Lsun$)
far-IR emission and large amounts of dust and molecular gas in a
number of high-$z$ quasars and radio galaxies (e.g.,
\citebare{Dunlopetal94};
\citebare{McMahonetal94}; \citebare{Ohtaetal96}; 
Omont {\it et al.}\ 1996a,b; \citebare{Guilloteauetal97}) 
are directly relevant in this
context. 
With their high far-IR luminosities, large
dust masses and large (inferred) molecular gas masses, these distant
objects seem to be the more extreme high-$z$ counterparts of the local
ULIGs. 
While the AGNs in the high-$z$ far-IR luminous
objects contribute an unknown fraction to the powering of the far-IR
luminosity, this argument only changes the conclusion qualitatively,
since it would be extremely surprising if intense star formation were
{\it not\/} taking place in such immense agglomerations of gas and
dust; furthermore, many local ULIGs also possess an AGN contributing
to the far-IR luminosity.  
Since dust emission has now been found in a
number of high-$z$ radio galaxies, such object may turn out to be
Rosetta stones for understanding the formation of massive galaxies.
\nocite{Omontetal96a} \nocite{Omontetal96b}

Given these results it is important to examine the view that
these far-IR luminous high-$z$ objects can indeed be regarded as 
forming by monolithic collapse. The expected parameters for such objects can be
estimated by considering a gas mass $\qu{M}{g}$ collapsing and forming
stars on a free fall timescale $\qu{t}{ff}$, resulting in a star
formation rate $\Mstardot=\qu{M}{g}/\qu{t}{ff}$. Starting with an
isothermal sphere, the resulting
star formation rate is
\begin{equation}
{\Mstardot\over {\rm M}_{\odot}\pun{yr}{-1}}=
40\,\left({\qu{v}{c}\over 100\kms}\right)^3,\label{eq.Mdot}
\end{equation}
where $\qu{v}{c}$ is the circular velocity,
which can be related to the enclosed mass $M$ within
radius $R$ using
\begin{equation}
\qu{v}{c}^2={GM\over R}.\label{eq.virialmass}
\end{equation}
The monolithic collapse scenario will be a valid approach if the
buildup of the initial starburst occurs on a timescale shorter than
that of merger-driven evolution. As shown by
\eqsref{eq.Mdot}{eq.virialmass}, this condition will be fulfilled
in the most massive systems, which have the highest star formation rates.
Indeed, star formation rates estimated for far-IR luminous high-$z$
galaxies are among the highest derived for any object in the universe,
consistent with the present approach of analyzing these objects as
galaxies forming by monolithic collapse.

\begin{figure}[!t]
\centerline{\psfig{figure=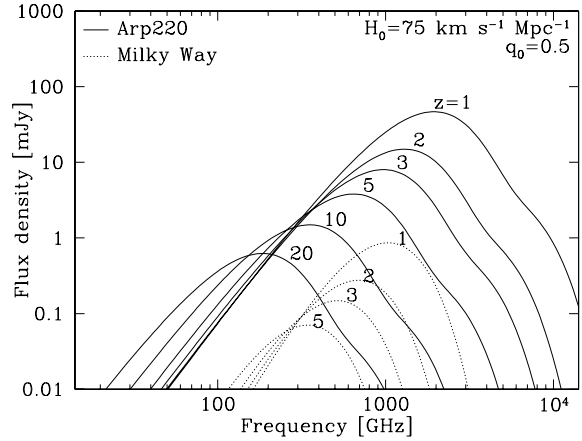,height=6cm,angle=270}}
\caption[]{\footnotesize Spectral energy distributions of the 
thermal dust emission of $\sou{Arp}{220}$ and the
Milky Way at various redshifts.\label{fig.Arpflux}}
\end{figure}

\section{Observations and implications}\label{sec.obs}

A dusty starburst galaxy emits more than 90\% of its luminosity in the
far-IR wavelength region.  The thermal dust emission obeys the
relation $S_\nu\propto Q_\nu B_\nu(\qu{T}{d})$, where
$B_\nu(\qu{T}{d})$ is the Planck function at the dust temperature
$\qu{T}{d}$, and $Q_\nu$ is the dust emissivity, which in the
submillimetre region obeys the relation $Q_\nu\propto\nu^{\beta}$, with $\beta$
increasing with wavelength from 1 to 2.  Therefore,
in the long wavelength Rayleigh-Jeans regime, the thermal dust
emission falls off as $S_\nu\propto\nu^{4}$. Hence for dusty high-$z$
objects, where the peak of the thermal dust emission is shifted into
the (sub)millimetre regime, a very large, negative $K$-correction
applies. As a result, the (sub)millimetre regime is the {\it ideal\/}
spectral region for studying dusty high-$z$ starburst galaxies. This
statement is illustrated by \figref{fig.Arpflux}, which shows that the
observed $850\mum$ flux of a galaxy with the far-IR spectral energy
distribution (SED) of $\sou{Arp}{220}$ does {\it not\/} decrease as
the redshift increases from 1 to 10. For the adopted cosmology,
$\sou{Arp}{220}$ would have an $850\mum$ flux density of about
$3\un{mJy}$ in this redshift range, a value within reach of the most
sensitive present-day submillimetre facilities such as the
Submillimetre Common User Bolometer Array (SCUBA) at the $15\un{m}$
James Clerk Maxwell Telescope (JCMT).  \Figref{fig.Arpflux}
furthermore shows that given a submillimetre detection of a galaxy of
unknown redshift, any redshift between 1 and 10 is equally likely.
Finally, it can be seen that even in the absence of an accurate
redshift determination, the far-IR luminosity can be reasonably well
estimated, provided the redshift is between 1 and 10. As shown by
\citetext{BlainLongair93}, a deep submillimetre-selected sample will
contain a very high proportion of high-$z$ galaxies. These properties
are unique to the submillimetre region and make this spectral interval
the ideal regime for studying star formation in the high-$z$ universe.

Before examining submillimetre observations of high-$z$ radio
galaxies, it is useful to analyse the assumptions and uncertainties in
the derivation of physical quantities from the observational results.
The quantities of most fundamental importance of the starformation
rate $\Mstardot$ and the molecular gas mass $M_{{\rm H}_2}$. Since the
latter quantity is sometimes derived from the dust mass $\qu{M}{d}$,
this quantity will also be discussed.

\begin{itemize}
\item
The star formation rate $\Mstardot$ is readily derived from the far-IR
luminosity $\qu{L}{FIR}$, using
\begin{equation}
{\Mstardot\over\Msun\,\pun{yr}{-1}} = 
\mbox{{\rm A}}\times10^{-10}\,{\qu{L}{FIR}\over\Lsun},
\end{equation}
where the constant A is of order unity and depends on the details of
the initial mass function.  The most important uncertainty in
this calculation stems from the assumption that the far-IR luminosity is
totally powered by star formation, which, given the presence of
powerful AGNs in high-$z$ radio galaxies, is a very dubious step.
Furthermore, in order to estimate $\qu{L}{FIR}$ from flux densities
determined on the Rayleigh-Jeans side of the SED, knowledge of the SED
is required. In the absence of observations at shorter (i.e., far-IR)
wavelengths, an intrinsic SED must be {\it assumed}, adding considerably to
the uncertainty in the derived star formation rate.
\item
The molecular gas mass $M_{{\rm H}_2}$ is normally derived from
emission in CO rotational lines, assuming a conversion factor that is
essentially an empirical conversion based on observations of Galactic
giant molecular clouds. An extensive literature exists on the general
validity of this conversion factor in other galaxies. The conversion
factor can be written either in terms of $\Htwo$ mass $\MHtwo$ through
\begin{equation}
{M_{{\rm H}_2}\over\Msun} = 
\alpha\,{L'_{\rm CO}\over{\rm K\,km\,s}^{-1}\,{\rm pc}^2},
\label{eq.MH2}
\end{equation}
where the derivation of $L'_{\rm CO}$ is discussed in
\citetext{VanDerWerfIsrael96b}, or in terms of $\Htwo$ column density
through
\begin{equation}
{N(\Htwo)\over{\rm cm}^{-2}} = 
\alpha'\,{\qu{I}{CO}\over{\rm K\,km\,s}^{-1}},
\label{eq.NH2}
\end{equation}
where $\qu{I}{CO}$ is the velocity-integrated CO
main beam brightness temperature in K$\kms$. The constants in
\eqsref{eq.MH2}{eq.NH2} are related through
$\alpha'=6.25\cdot10^{19}\,\alpha$, and are proportional to
$\sqrt{n_{{\rm H}_2}}/\qu{T}{b}$, where $n_{{\rm H}_2}$ is the $\Htwo$
number density and $\qu{T}{b}$ is the intrinsic brightness temperature
of the relevant CO transition (see e.g., \citebare{Maloney90} for a simple
derivation of this result). Given the dependence on
physical conditions, it is very uncertain whether the standard
Galactic conversion factor $\alpha = 4$ (equivalent to
$\alpha'=2.5\cdot10^{20}$) is valid in such extreme objects as
dusty high-$z$ radio galaxies. In ULIGs, the molecular gas is both
denser and hotter than in the Milky Way. Although these effects tend
to cancel eachother, the actual value of $\alpha$ in these objects
remains uncertain. Detailed studies by \citetext{BryantScoville96} and
\citetext{Solomonetal97} indicate that in ULIGs, $\alpha$ may be up
to a factor of 4 lower than in Galactic molecular clouds.
\item
The molecular gas mass can also estimated from the dust mass assuming
a ratio of gas mass to dust mass, defined by
$X=\qu{M}{g}/\qu{M}{d}$. The approriate value for high-$z$ radio
galaxies is difficult to choose {\it a priori}, given that $X$ ranges
from $\la150$ in the Milky Way \cite{SavageMathis79},
through approximately 500 in nearby spirals \cite{DevereuxYoung90} and
low-$z$ ULIGs \cite{Sandersetal91}, to much higher values in damped
$\Lya$ systems at $z>2$ \cite{Falletal89,Pettinietal94}. However, in
the four high-$z$ QSOs where both CO and dust emission have been
detected (H1413+117, $\sou{IRAS}{F10241{+}4724}$,
$\sou{BR}{1202{-}0725}$ and $\sou{BRI}{1335{-}0417}$), the data point to
a value of $X\sim500$, again similar to that in local ULIGs.\\
The dust mass $\qu{M}{d}$ follows from
\begin{equation}
\qu{M}{d}={1\over 1+z}\,{S_{\nu}\,\qu{D}{L}^2\over \qu{k}{d}(\nu_0)\,B_{\nu_0}(\qu{T}{d})}\label{eq.Md},
\end{equation}
where $S_{\nu}$ is the thermal flux density observed at frequency
$\nu$, $\qu{k}{d}(\nu_0)$ is the dust mass absorption coefficient at
rest frequency $\nu_0=\nu(1+z)$, $\qu{T}{d}$ is the temperature of the
dust grains, $B_{\nu}(T)$ is the Planck function at frequency $\nu$
and temperature $T$, and $\qu{D}{L}$ is the
luminosity distance. In the Rayleigh-Jeans regime, $\qu{T}{d}$ enters
in \eqref{eq.Md} in the first power, and given that its plausible
range is from 30 to $70\un{K}$, it produces some uncertainty in the
derived dust mass, that can only be reduced by observing a
well-sampled SED, including the temperature-sensitive Wien
side. However, $\qu{k}{d}$ is a more serious source of
uncertainty. Estimates of $\qu{k}{d}$ at $800\mum$ in the Milky Way
cover almost two orders of magnitude (e.g., \citebare{DraineLee84}
versus \citebare{RowanRobinson86}). As a reasonable compromise, a value
$\qu{k}{d}=1\pun{cm}{2}\pun{g}{-1}$ at $800\mum$ is adopted here, and
$\qu{k}{d}\propto\nu^2$ for $\lambda>100\mum$,
which leaves about a factor of 2.5
uncertainty in both directions. Finally, a more subtle uncertainty
arises from the implicit assumption of isothermal dust emission, the
dangers of which are demonstrated by \citetext{Draine90a} in an
instructive example.
\end{itemize}

\begin{table}[!t]
{\footnotesize
\begin{center}
\caption{\footnotesize Submillimetre observations of dust emission in
high-$z$ radio galaxies. Upper limits represent $3\sigma$ values and
are only included for observations carried out with
bolometer array instruments. In
order to derive far-IR luminosities, an $\sou{Arp}{220}$-like SED has
been assumed. Dust masses are calulated for a dust temperature of
$50\un{K}$, appropriate for the adopted SED\null.}
\label{tab.results}
\begin{tabular*}{\hsize}{@{\extracolsep\fill}lcrrrc}
\hline\\[-3mm]
Name & $z$ & $S_{\rm 0.85\,mm}$ & $\qu{L}{FIR}$ & $\qu{M}{dust}$ & notes\\
 & & [mJy] & [$\Lsun$] & [$\Msun$] & \\
\hline\\[-3mm]
$\sou{6C}{0140{+}326}$  & 4.41 & $<4.5$ & $<6.8\cdot10^{12}$ & $<1.5\cdot10^8$ & 1 \\
$\sou{8C}{1435{+}635}$  & 4.25 & $8.3$  & $1.2\cdot10^{13}$  & $2.8\cdot10^8$  & 2 \\
$\sou{4C}{41.17}$       & 3.80 & $12.3$ & $1.7\cdot10^{13}$  & $4.3\cdot10^8$  & 3 \\
$\sou{4C}{60.07}$       & 3.79 & $11.0$ & $1.5\cdot10^{13}$  & $3.8\cdot10^8$  & 1 \\
$\sou{6C}{0032{+}412}$  & 3.67 & $<4.8$ & $<6.6\cdot10^{12}$ & $<1.7\cdot10^8$ & 3 \\
$\sou{6C}{1909{+}722}$  & 3.54 & $12.4$ & $1.5\cdot10^{13}$  & $4.4\cdot10^8$  & 1 \\
$\sou{6C}{1232{+}39}$   & 3.22 & $3.9$  & $5.0\cdot10^{12}$  & $1.4\cdot10^8$  & 3 \\
$\sou{MG}{1019{+}0535}$ & 2.76 & 14.7   & $1.8\cdot10^{13}$  & $5.7\cdot10^8$  & 4 \\
$\sou{3C}{257}$         & 2.47 & 3.9    & $4.5\cdot10^{12}$  & $1.6\cdot10^8$  & 3 \\
$\sou{53W}{002}$        & 2.39 & $<3.3$ & $<3.7\cdot10^{12}$ & $<1.3\cdot10^8$ & 3 \\
$\sou{6C}{0930{+}38}$   & 2.39 & $<3.6$ & $<4.0\cdot10^{12}$ & $<1.4\cdot10^8$ & 3 \\
$\sou{6C}{0901{+}35}$   & 1.91 & $<3.6$ & $<3.7\cdot10^{12}$ & $<1.5\cdot10^8$ & 3 \\
$\sou{6C}{0905{+}39}$   & 1.88 & 2.8    & $2.8\cdot10^{12}$  & $1.2\cdot10^8$  & 3 \\
$\sou{6C}{1204{+}37}$   & 1.78 & $<3.0$ & $<2.9\cdot10^{12}$ & $<1.3\cdot10^8$ & 3 \\
$\sou{3C}{294}$         & 1.78 & $<2.4$ & $<2.3\cdot10^{12}$ & $<1.0\cdot10^8$ & 3 \\
$\sou{3C}{241}$         & 1.62 & $<4.5$ & $<4.2\cdot10^{12}$ & $<2.0\cdot10^8$ & 3 \\
$\sou{3C}{324}$         & 1.21 & $<3.1$ & $<2.4\cdot10^{12}$ & $<1.4\cdot10^8$ & 3,5 \\
$\sou{3C}{217}$         & 0.89 & $<2.4$ & $<1.5\cdot10^{12}$ & $<9.9\cdot10^7$ & 3 \\
$\sou{3C}{265}$         & 0.81 & $<3.0$ & $<1.7\cdot10^{12}$ & $<1.2\cdot10^8$ & 3 \\
$\sou{3C}{340}$         & 0.78 & $<3.6$ & $<2.0\cdot10^{12}$ & $<1.4\cdot10^8$ & 3 \\
$\sou{3C}{277.2}$       & 0.76 & $<3.3$ & $<1.8\cdot10^{12}$ & $<1.3\cdot10^8$ & 3 \\
\hline\\[-3mm]
\noalign{\leftline{$^1$ unpublished SCUBA data\ \ $^2$
    \protect\citetext{Ivisonetal98}\ \ 
$^3$ \protect\citetext{HughesDunlop98}}}
\noalign{\leftline{$^4$ \protect\citetext{Cimattietal98}\ \ 
$^5$ \protect\citetext{Bestetal98b}}}\\[-10mm]
\end{tabular*}
\end{center}}
\end{table}

\Tabref{tab.results} summarizes the current observational situation in
submillimetre studies of radio galaxies at cosmologically significant
redshifts. Given the uncertainties described above it is
advisable to minimize the number of uncertain steps in the
analysis. Hence we concentrate on the derived far-IR luminosities and
dust masses, and avoid the additional uncertainties resulting from
converting these values into star formation rates and molecular gas masses.

Inspection of \tabref{tab.results} shows that the success rate for
the detection of thermal submillimetre emission from radio galaxies
is highest for high-$z$ objects: 
for $z>2.4$, the detection rate is 77\%, while for
$z<2.4$ the detection rate is only 8\%. It is tempting to interpret
this result as an evolution effect in the radio galaxy population.
In this case it would be concluded that 
large far-IR luminosities and dust masses in radio
galaxies are much more common at high redshift than at low
redshift. Interpreting the far-IR luminosities as reflecting immense
star formation rates in extremely gas-rich galaxies (as concluded
from the large dust masses), these results provide strong support for
the ``initial starburst'' scenario. Indeed, derived star formation
rates are in the range suggested by \eqref{eq.Mdot}. Taking 
$\sou{4C}{41.17}$ as an example, the inferred star formation rate
would be $\sim1700\Msun\pun{yr}{-1}$. This value is significantly
higher than the value $\Mstardot\sim60-440\Msun\pun{yr}{-1}$ derived
by \citetext{Deyetal97} from spectroscopy of the young stellar
population in the rest-frame ultraviolet, but given the large dust mass in
$\sou{4C}{41.17}$, significant extinction is expected, naturally
accounting for the discrepancy. Assuming a gas/dust mass ratio of 500,
the far-IR derived star formation rate would turn the entire gas
reservoir into stars in a burst lasting about $10^8\un{yr}$, which is
a typical time scale for a starburst.

However, there is a subtle selection effect that may be responsible
for the high success rate of the detection of thermal submillimetre
emission for $z>2.4$. The highest redshift objects in
\tabref{tab.results} also have the highest radio powers. Given the
good correlation between radio power and far-IR luminosity in low-$z$
radio galaxies (e.g., \citebare{Heckmanetal94}; \citebare{Hesetal95}),
it is to be expected that the highest redshift radio galaxies in
\tabref{tab.results} also have the highest far-IR luminosities. But
since (as shown in \figref{fig.Arpflux}) the observed submillimetre
flux density does not drop very much with redshift for a given far-IR
luminosity, this situation will naturally lead to a higher
submillimetre detection rate for the higher redshift galaxies, simply
because the way they are selected makes them intrinsically more
luminous. The high submillimetre detection rate for $z>2.4$ radio
galaxies may result entirely from this effect. Perhaps significantly,
one of the two high-$z$ radio galaxies not detected in the
submillimetre, $\sou{6C}{0140{+}326}$ at $z=4.41$, has a much lower
radio power than the other high-$z$ radio galaxies in
\tabref{tab.results} \cite{Rawlingsetal95}.  Thus, before a
conclusion can be reached on the role of dust and far-IR emission in
the evolution of the radio galaxy population, this selection effect
must be addressed. Submillimetre observations of intrinsically fainter
high-$z$ radio sources will be able to quantify the role of radio
power in the trends signalled above.  Intrinsically fainter samples of
high-$z$ radio galaxies are now becoming available (e.g.,
\citebare{Ealesetal97}), and SCUBA observations of these sources will
be able to supply the required information. This is an extremely
urgent, and entirely feasible experiment.

\section{Radio galaxies as luminous infrared galaxies\label{sec.local}}

The dangers of blindly using the conversion factors given in
\secref{sec.obs} for calculating molecular gas masses are illustrated
by considering molecular gas mass estimates for $\sou{4C}{41.17}$. Using a
gas/dust ratio of 500, an $\Htwo$ mass
$\MHtwo\sim2\times10^{11}\Msun$ is derived. Using \eqref{eq.MH2}, the
corresponding CO signal in the low-$J$ lines is predicted to be
$\qu{L}{CO}'\sim5\times10^{10}\un{K}\kms\pun{pc}{2}$. However, the
observed $3\sigma$ upper limit is
$3\times10^{10}\un{K}\kms\pun{pc}{2}$ \cite{Evansetal96}. Allowing
$\alpha$ to become smaller than 4 in \eqref{eq.MH2} makes the
discrepancy larger. Evidently, at least one of the conversion steps
used in this analysis is in error. 

\begin{figure}[!t]
\centerline{\psfig{figure=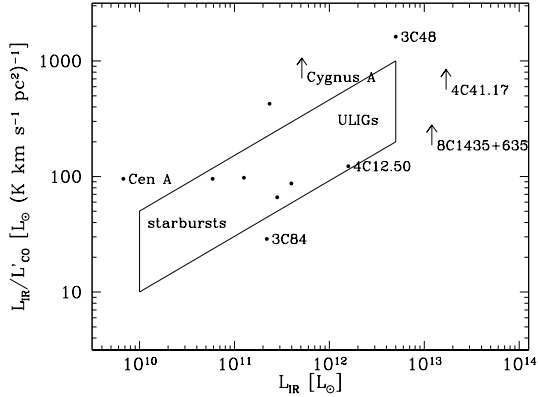,height=5.5cm,angle=270}}
\caption[]{\footnotesize Comparison of far-IR and CO luminosities of
  of low-$z$ and high-$z$ radio galaxies. The region occupied by local
  starburst galaxies and ULIGs (from \citebare{SandersMirabel96}) is
  indicated. Unlabeled point represent detections of compact and FR
  class I
  low-$z$ radio galaxies from \citetext{Mazzarellaetal93}.
  \label{fig.FIRCO}}
\end{figure}

\begin{table}[!t]
{\footnotesize
\begin{center}
\caption{\footnotesize Parameters of
low-redshift radio galaxies and radio-loud QSOs detected in the
far-infrared and (except Cygnus~A) 
in CO emission. $\Htwo$ masses have been derived
according to \eqref{eq.MH2}. Upper limits are $3\sigma$ values.}
\label{tab.lowz}
\begin{tabular*}{\hsize}{@{\extracolsep\fill}lccccc}
\hline\\[-3mm]
Name & $z$ & $\qu{L}{FIR}$ & $\qu{L}{CO}'$ &$\MHtwo$ & notes\\
 & & [$\Lsun$] & [K$\kms\pun{pc}{2}$] & [$\Msun$] & \\
\hline\\[-3mm]
$\sou{3C}{48}$    & 0.369 & $5.0\cdot10^{12}$ & $3.1\cdot10^9$    & $1.2\cdot10^{10}$ & 1 \\
$\sou{4C}{12.50}$ & 0.122 & $1.6\cdot10^{12}$ & $1.3\cdot10^{10}$ & $5.2\cdot10^{10}$ & 2 \\
Cygnus~A          & 0.057 & $5.1\cdot10^{11}$ & $<7.2\cdot10^8$   & $<2.9\cdot10^9$ & 3 \\
$\sou{3C}{84}$    & 0.018 & $2.2\cdot10^{11}$ & $7.6\cdot10^9$    & $3.0\cdot10^{10}$ & 4 \\
Centaurus~A       & 0.002 & $6.8\cdot10^{9}$  & $7.2\cdot10^7$    & $2.9\cdot10^8$ & 5 \\
\hline\\[-3mm]
\noalign{\leftline{$^1$ \protect\citetext{Winketal97}\ \ 
$^2$ \protect\citetext{Mirabeletal89}\ \ $^3$ \protect\citetext{Mazzarellaetal93}}}
\noalign{\leftline{$^4$ \protect\citetext{Inoueetal96}\ \ 
$^5$ \protect\citetext{Eckartetal90}}}\\[-10mm]
\end{tabular*}
\end{center}}
\end{table}

This example shows the importance
of minimizing the number of uncertain steps, and the need to analyse
the results in terms of direct observables. For a proper analysis of
the CO upper limits and submillimetre detections in radio galaxies,
it is most useful to 
plot $\qu{L}{FIR}/\qu{L}{CO}'$ as a function of $\qu{L}{FIR}$, as in
\figref{fig.FIRCO}. This diagram includes some data on low-$z$ radio
galaxies and radio-loud QSOs detected in the far-IR and in CO,
summarized in \tabref{tab.lowz}. \Figref{fig.FIRCO} shows that 
the ratio $\qu{L}{FIR}/L'_{\rm CO}$
in local star forming galaxies is strongly luminosity
dependent. Low-$z$ radio galaxies cover a similar range of values, but
there are significant exceptions, such as
Cygnus~A\null. \Figref{fig.FIRCO} allows an assessment of the
significance of the CO non-detections in the dusty high-$z$ radio
galaxies
$\sou{4C}{41.17}$ \cite{Evansetal96} and $\sou{8C}{1435{+}635}$ 
\cite{Ivisonetal98}. It is
evident that these upper limits still allow parameters consistent with local
ULIGs, and are not significant, given the luminosity dependence of the ratio
$\qu{L}{FIR}/L'_{\rm CO}$. However, current
millimetre wave interferometers are capable of significantly deeper
measurements, and should be able to detect CO emission from the most
luminous dusty high-$z$ radio galaxies in the coming years. 

For systems where the far-IR luminosity is mostly powered by star formation,
\figref{fig.FIRCO} shows that the {\it star formation efficiency\/}
(i.e., star formation rate per unit molecular gas mass, as measured by
the ratio $\qu{L}{FIR}/L'_{\rm CO}$) increases with star formation rate
(as measured by $\qu{L}{FIR}$). Thus, if the far-IR luminosity of
dusty high-$z$ radio galaxies is powered by star formation, star
formation in these objects proceeds extremely efficiently, reinforcing
the view of an ``initial starburst'' rapidly converting most of the
available gas mass into stars, as in monolithic collapse. It is
therefore extremely urgent to obtain CO detections of these high-$z$
radio galaxies, so that their location in \figref{fig.FIRCO} can be
determined. Subsequently the question whether star formation dominates
the far-IR luminosity must be addressed. A crucial parameter here is
the spatial extent of the submillimetre emission. Extended emission
would indicate a distributed source of heating, strongly supporting
the view that star formation is responsible for the observed far-IR
luminosity. A future large (sub)millimetre array will be able to
address this question observationally.

\begin{acknow}
I would like to thank my collaborators Philip Best and
Huub R\"ottgering for various
dicussions on this subject, and Remo Tilanus for his excellent support
at the James Clerk Maxwell Telescope.
\end{acknow}

\bibliographystyle{astrobib_knaw}
\bibliography{%
strings,%
dampedLya,%
dust,%
galaxyevolution,%
galaxyformation,%
mm,%
radiogalaxies,%
BR1202-0725,%
BRI1335-0417,%
3C48,%
4C41.17,%
6C0140+326,%
8C1435+635,%
CenA,%
ISMgalaxies,%
Mrk231,%
NGC1275,%
ULIGs}

\begin{references}
\astrobibitem{Baugh {\it et al.}}{1998}{Baughetal98}
Baugh, C.M., Cole, S., Frenk, C.S., \& Lacey, C.G.\ 1998, ApJ, 498, 504

\astrobibitem{Best {\it et al.}}{1998a}{Bestetal98a}
Best, P.N., Longair, M.S., \& R{\"o}ttgering, H.J.A., 1998a, MNRAS, in press

\astrobibitem{Best {\it et al.}}{1998b}{Bestetal98b}
Best, P.N., R{\"o}ttgering, H.J.A., Bremer, M.N., Cimatti, A., Miley, G.K.,
  Pentericci, L., Tilanus, R.P.J., \& Van~der Werf, P.P., 1998b, in preparation

\astrobibitem{Blain \& Longair}{1993}{BlainLongair93}
Blain, A.W., \& Longair, M.S.\ 1993, MNRAS, 264, 509

\astrobibitem{Bryant \& Scoville}{1996}{BryantScoville96}
Bryant, P.M., \& Scoville, N.Z.\ 1996, ApJ, 678, 457

\astrobibitem{Cimatti {\it et al.}}{1998}{Cimattietal98}
Cimatti, A., Freudling, W., R{\"o}ttgering, H.J.A., Ivison, R.J., \& Mazzei,
  P.\ 1998, A\&A, 329, 399

\astrobibitem{Devereux \& Young}{1990}{DevereuxYoung90}
Devereux, N.A., \& Young, J.S.\ 1990, ApJ, 359, 42

\astrobibitem{Dey {\it et al.}}{1997}{Deyetal97}
Dey, A., Van Breugel, W., Vacca, W.D., \& Antonucci, R.\ 1997, ApJ, 490, 698

\astrobibitem{Draine \& Lee}{1984}{DraineLee84}
Draine, B.T., \& Lee, H.M.\ 1984, ApJ, 285, 89

\astrobibitem{Draine}{1990}{Draine90a}
Draine, B.T., 1990, Mass determinations from far-infrared observations. In:\
  Thronson, H.A., \& Shull, J.M.\ (eds.), The interstellar medium in galaxies,
  (Dordrecht: Kluwer), p.~483

\astrobibitem{Dunlop {\it et al.}}{1994}{Dunlopetal94}
Dunlop, J.S., Hughes, D.H., Rawlings, S., Eales, S.A., \& Ward, M.J.\ 1994,
  Nat, 370, 347

\astrobibitem{Eales {\it et al.}}{1997}{Ealesetal97}
Eales, S., Rawlings, S., Law-Green, D., Cotter, G., \& Lacy, M.\ 1997, MNRAS,
  291, 593

\astrobibitem{Eckart {\it et al.}}{1990}{Eckartetal90}
Eckart, A., Cameron, M., Rothermel, H., Wild, W., Zinnecker, H., Rydbeck, G.,
  Olberg, M., \& Wiklind, T.\ 1990, ApJ, 363, 451

\astrobibitem{Evans {\it et al.}}{1996}{Evansetal96}
Evans, A.S., Sanders, D.B., Mazzarella, J.M., Solomon, P.M., Downes, D.,
  Kramer, C., \& Radford, S.J.E.\ 1996, ApJ, 457, 658

\astrobibitem{Fall {\it et al.}}{1989}{Falletal89}
Fall, S.M., Pei, Y.C., \& McMahon, R.G.\ 1989, ApJ, 337, 7

\astrobibitem{Guilloteau {\it et al.}}{1997}{Guilloteauetal97}
Guilloteau, S., Omont, A., McMahon, R.G., Cox, P., \& Petitjean, P.\ 1997,
  A\&A, 328, L1

\astrobibitem{Heckman {\it et al.}}{1994}{Heckmanetal94}
Heckman, T.M., O'Dea, C.P., Baum, S.A., \& Laurikainen, E.\ 1994, ApJ, 428, 65

\astrobibitem{Hes {\it et al.}}{1995}{Hesetal95}
Hes, R., Barthel, P.D., \& Hoekstra, H.\ 1995, A\&A, 303, 8

\astrobibitem{Hughes \& Dunlop}{1998}{HughesDunlop98}
Hughes, D.H., \& Dunlop, J.S., 1998, to appear in {Highly Redshifted Radio
  Lines}, eds. {C. L. Carilli, S. J. E. Radford, K. Menten, \& G. Langston, ASP
  Conference Series}

\astrobibitem{Inoue {\it et al.}}{1996}{Inoueetal96}
Inoue, M.Y., Kameno, S., Kawabe, R., Inoue, M., Hasegawa, T., \& Tanaka, M.\
  1996, AJ, 1996, 1853

\astrobibitem{Ivison {\it et al.}}{1998}{Ivisonetal98}
Ivison, R.J., et al.\ 1998, ApJ, 494, 211

\astrobibitem{Kauffmann {\it et al.}}{1993}{Kauffmannetal93}
Kauffmann, G., White, S.D.M., \& Guideroni, B.\ 1993, MNRAS, 261, 921

\astrobibitem{Maloney}{1990}{Maloney90}
Maloney, P., 1990, Mass determinations from {CO} observations. In:\ Thronson,
  H.A., \& Shull, J.M.\ (eds.), The interstellar medium in galaxies,
  (Dordrecht: Kluwer), p.~493

\astrobibitem{Matthews {\it et al.}}{1964}{Matthewsetal64}
Matthews, T.A., Morgan, W.W., \& Schmidt, M.\ 1964, ApJ, 140, 35

\astrobibitem{Mazzarella {\it et al.}}{1993}{Mazzarellaetal93}
Mazzarella, J.M., Graham, J.R., Sanders, D.B., \& Djorgovski, S.\ 1993, ApJ,
  409, 170

\astrobibitem{McMahon {\it et al.}}{1994}{McMahonetal94}
McMahon, R.G., Omont, A., Bergeron, J., Kreysa, E., \& Haslam, C.G.T.\ 1994,
  MNRAS, 267, L9

\astrobibitem{Miley {\it et al.}}{1992}{Mileyetal92}
Miley, G.K., Chambers, K.C., Van Breugel, W.J.M., \& Macchetto, F.\ 1992, ApJ,
  401, L69

\astrobibitem{Mirabel {\it et al.}}{1989}{Mirabeletal89}
Mirabel, I.F., Sanders, D.B., \& Kaz{\`e}s, I.\ 1989, ApJ, 340, L9

\astrobibitem{Ohta {\it et al.}}{1996}{Ohtaetal96}
Ohta, K., Yamada, T., Nakanishi, K., Kohno, K., Akiyama, M., \& Kawabe, R.\
  1996, Nat, 382, 426

\astrobibitem{Omont {\it et al.}}{1996a}{Omontetal96b}
Omont, A., McMahon, R.G., Cox, P., Kreysa, E., Bergeron, J., Pajot, F., \&
  Storrie-Lombardi, L.J.\ 1996a, A\&A, 315, 1

\astrobibitem{Omont {\it et al.}}{1996b}{Omontetal96a}
Omont, A., Petitjean, P., Guilloteau, S., McMahon, R.G., Solomon, P.M., \&
  Pecontal, E.\ 1996b, Nat, 382, 428

\astrobibitem{Pentericci {\it et al.}}{1997}{Pentericcietal97}
Pentericci, L., R{\"ot}tgering, H.J.A., Miley, G.K., Carilli, C.L., \&
  McCarthy, P.\ 1997, A\&A, 326, 580

\astrobibitem{Pettini {\it et al.}}{1994}{Pettinietal94}
Pettini, M., Smith, L.J., Hunstead, R.W., \& King, D.L.\ 1994, ApJ, 426, 79

\astrobibitem{Rawlings {\it et al.}}{1995}{Rawlingsetal95}
Rawlings, S., Lacy, M., Blundell, K.M., Eales, S.A., Bunker, A.J., \&
  Garrington, S.T.\ 1995, Nat, 383, 502

\astrobibitem{Rowan-Robinson}{1986}{RowanRobinson86}
Rowan-Robinson, M.\ 1986, MNRAS, 219, 737

\astrobibitem{Sanders \& Mirabel}{1996}{SandersMirabel96}
Sanders, D.B., \& Mirabel, I.F.\ 1996, ARA\&A, 34, 749

\astrobibitem{Sanders {\it et al.}}{1991}{Sandersetal91}
Sanders, D.B., Scoville, N.Z., \& Soifer, B.T.\ 1991, ApJ, 370, 158

\astrobibitem{Savage \& Mathis}{1979}{SavageMathis79}
Savage, B.D., \& Mathis, J.S.\ 1979, ARA\&A, 17, 73

\astrobibitem{Solomon {\it et al.}}{1997}{Solomonetal97}
Solomon, P.M., Downes, D., Radford, S.J.E., \& Barrett, J.W.\ 1997, ApJ, 478,
  144

\astrobibitem{Van~der Werf \& Israel}{1996}{VanDerWerfIsrael96b}
Van~der Werf, P.P., \& Israel, F.P., 1996, Studying high redshift starburst
  galaxies with a large (sub)millimetre array. In:\ Shaver, P.A.\ (ed.),
  Science with large millimetre arrays, {ESO Astrophyscis Symposia,} (Berlin:
  Springer), p.~51

\astrobibitem{White \& Frenk}{1991}{WhiteFrenk91}
White, S.D.M., \& Frenk, C.S.\ 1991, ApJ, 379, 52

\astrobibitem{Wink {\it et al.}}{1997}{Winketal97}
Wink, J.E., Guilloteau, S., \& Wilson, T.L.\ 1997, A\&A, 322, 427

\end{references}

\end{document}